\documentclass[twocolumn,secnumarabic,amssymb,amsmath,nofootinbib,tightenlines,nobibnotes,aps,prl,showpacs]{revtex4}
\usepackage{graphicx,epsfig,color}
\begin{document}

\def\cagl{$^{1}$}
\def\cern{$^{2}$}
\def\lpc{$^{3}$}
\def\heid{$^{4}$}
\def\lis{$^{5}$}
\def\llr{$^{6}$}
\def\riken{$^{7}$}
\def\suny{$^{8}$}
\def\torino{$^{9}$}
\def\ipnl{$^{10}$}
\def\yer{$^{11}$}

\title{First Measurement of the $\rho$ Spectral Function in High-Energy Nuclear Collisions}
\author{
R.~Arnaldi$^{9}$, 
R.~Averbeck$^{8}$, 
K.~Banicz$^{4,2}$, 
J.~Castor$^{3}$, 
B.~Chaurand$^{6}$, 
C.~Cical\`o$^{1}$, 
A.~Colla$^{9}$, 
P.~Cortese$^{9}$, 
S.~Damjanovic$^{4}$, 
A.~David$^{5,2}$, 
A.~De~Falco$^{1}$, 
A.~Devaux$^{3}$, 
A. Drees$^{8}$, 
L.~Ducroux$^{10}$, 
H.~En'yo$^{7}$,
J.~Fargeix$^{3}$,  
A.~Ferretti$^{9}$, 
M.~Floris$^{1}$,
A.~F\"{o}rster$^{2}$,  
P.~Force$^{3}$, 
N.~Guettet$^{2,3}$, 
A.~Guichard$^{10}$, 
H.~Gulkanian$^{11}$, 
J.~M.~Heuser$^{7}$, 
M.~Keil$^{5,2}$, 
L.~Kluberg$^{6,2}$,
C.~Louren\c{c}o$^{2}$,
J.~Lozano$^{5}$, 
F.~Manso$^{3}$, 
A.~Masoni$^{1}$, 
P.~Martins$^{5,2}$, 
A.~Neves$^{5}$, 
H.~Ohnishi$^{7}$, 
C.~Oppedisano$^{9}$, 
P.~Parracho$^{2}$, 
Ph.~Pillot$^{10}$, 
G.~Puddu$^{1}$, 
E.~Radermacher$^{2}$, 
P.~Ramalhete$^{2}$, 
P.~Rosinsky$^{2}$, 
E.~Scomparin$^{9}$, 
J.~Seixas$^{5,2}$, 
S.~Serci$^{1}$, 
R.~Shahoyan$^{5,2}$, 
P.~Sonderegger$^{5}$, 
H.J.~Specht$^{4,2}$, 
R.~Tieulent$^{10}$, 
G.~Usai$^{1}$, 
R.~Veenhof$^{5,2}$, 
H.K.~W\"ohri$^{5,2}$ \\ 
(NA60 Collaboration)
}

\affiliation{
\cagl \mbox{Universit\`a di Cagliari and INFN, Cagliari, Italy}\\
\cern \mbox{CERN, Geneva, Switzerland}\\
\lpc \mbox{LPC, Universit\'e Blaise Pascal and CNRS-IN2P3, Clermont-Ferrand, France}\\
\heid \mbox{Universit\"{a}t Heidelberg, Heidelberg, Germany}\\
\lis \mbox{CFTP, Instituto Superior T\'ecnico, Lisbon, Portugal}\\
\llr \mbox{LLR, Ecole Polytechnique and CNRS-IN2P3, Palaiseau, France} \\
\riken \mbox{RIKEN, Wako, Saitama, Japan}\\
\suny \mbox{SUNY, Stony Brook, NY, USA}\\
\torino \mbox{Universit\`a di Torino and INFN, Turin, Italy}\\
\ipnl \mbox{IPNL, Universit\'e Claude Bernard Lyon-I and CNRS-IN2P3, Villeurbanne, France}\\
\yer \mbox{YerPhI, Yerevan, Armenia}\\
}

\begin{abstract}
We report on a precision measurement of low-mass muon pairs in
158 AGeV indium-indium collisions at the CERN SPS. A significant excess of
pairs is observed above the yield expected from neutral meson
decays. The unprecedented sample size of 360\,000 dimuons and the good
mass resolution of about 2\% allow us to isolate the excess by
subtraction of the decay sources. The shape of the resulting mass
spectrum is consistent with a dominant contribution from
$\pi^{+}\pi^{-}\rightarrow\rho\rightarrow\mu^{+}\mu^{-}$
annihilation. The associated space-time averaged $\rho$ spectral function shows a strong
broadening, but essentially no shift in mass. This may rule out
theoretical models linking hadron masses directly to the chiral~condensate.
\end{abstract}

%\keywords{Heavy Ion Collisions, Lepton Pairs}%Use showkeys class
\maketitle

%%%introduction
According to quantum chromodynamics (QCD), strongly interacting matter
will, at sufficiently high temperatures or baryon densities, undergo a
phase transition from a state of hadronic constituents to quark
matter, a plasma of deconfined quarks and gluons. At the same time,
chiral symmetry, spontaneously broken in the hadronic world, will be
restored. High-energy nucleus-nucleus collisions provide the only way
to investigate this issue in the laboratory.

Among the observables used for the diagnostics of the hot and
dense fireball formed in these collisions, lepton pairs are
particularly attractive. In contrast to hadrons, they directly probe
the entire space-time evolution of the fireball and freely escape from
the interaction zone, undisturbed by final-state interactions. In the
hadronic phase, thermal dilepton production in the mass region $<$1
GeV/c$^{2}$ is largely mediated by the light vector mesons $\rho$, $\omega$
and $\phi$. Among these, the $\rho$(770 MeV/c$^{2}$) is the most important,
due to its strong coupling to the $\pi\pi$ channel and its 
lifetime of only 1.3 fm/c, much shorter than the lifetime of the
fireball. These properties have given it a key role as {\it the} test
particle for ``in-medium modifications'' of hadron properties close to
the QCD phase boundary. The ``dual'' role of the $\rho$, probing the
medium and having the medium probe its own properties, was pointed
out already more than 20 years ago: changes both in width and in
mass were suggested as precursor signatures of the chiral
transition~\cite{Pisarski:mq}. There seems to be some consensus now
that the {\it width} of the $\rho$ should increase towards the
transition region, based on a number of quite different theoretical
approaches~\cite{Pisarski:mq,Dominguez:1992dw,Pisarski:1995xu,Rapp:1995zy,Rapp:1999ej}.
On the other hand, no consensus exists on how the {\it mass} of the
$\rho$ should change in approaching the transition: predictions exist
for a decrease~\cite{Pisarski:mq, Brown:kk,Brown:2001nh,
Hatsuda:1991ez}, a constant behavior~\cite{Rapp:1995zy,Rapp:1999ej},
and even an increase~\cite{Pisarski:1995xu}. Strongly increasing widths
and nearly constant masses are also predicted for pions and for
nucleons~\cite{Leutwyler:1990uq}.

Experimentally, low-mass electron pair production was previously
investigated at the CERN SPS by the CERES/NA45 experiment for p-Be/Au,
S-Au and Pb-Au
collisions~\cite{Agakichiev:mv,Agakichiev:1995xb,Agakichiev:1997au}. The
common feature of all results from nuclear collisions was an excess of
the observed dilepton yield above the expected electromagnetic decays
of neutral mesons, by a factor of 2-3, for masses above 0.2~GeV/c$^{2}$. The
surplus yield has generally been attributed to direct thermal
radiation from the fireball, dominated by pion annihilation
$\pi^{+}\pi^{-}\rightarrow\rho\rightarrow l^{+}l^{-}$ with an
intermediate $\rho$ which is strongly modified by the
medium. Statistical accuracy and mass resolution of the data were,
however, not sufficient to reach any sensitivity to the {\it
character} of the in-medium changes.

In this Letter, we present first results on low-mass muon pair production from
the NA60 experiment at the CERN SPS. Compared to the results
from CERES, we have been able to improve the statistical accuracy by a
factor of $>$1000 and the mass resolution by a factor of 2-3. This has
allowed us, for the first time in this field, to isolate the excess by
subtraction of the expected sources. The essential features of the
resulting mass spectrum, interpreted as the space-time averaged $\rho$
spectral function associated with pion annihilation, can be summarized
in a few words: the $\rho$ strongly broadens, but does not show any
shift in mass.

The NA60 apparatus complements the muon spectrometer previously used
by NA50 with a high-granularity silicon pixel telescope of
unprecedented radiation tolerance~\cite{Gluca:2005,Keil:2005zq}. The
telescope, embedded in a 2.5~T dipole magnet in the vertex region,
tracks all charged particles before the hadron absorber and determines
their momenta independently of the muon spectrometer. The matching of
the muon tracks before and after the hadron absorber, both in {\it
angular and momentum} space, improves the dimuon mass resolution in
the region of the vector mesons $\omega$, $\phi$ from $\sim$80 to
$\sim$20~MeV/c$^{2}$, significantly reduces the combinatorial background due
to $\pi$ and K decays and makes it possible to measure the muon
offset with respect to the interaction vertex~\cite{Ruben:2005qm}. The
additional bend by the dipole field in the target region deflects
soft muons into the acceptance of the muon spectrometer,
thereby strongly enhancing the opposite-sign dimuon acceptance at low
mass and low transverse momentum with respect to all previous dimuon
experiments~\cite{michele:2005}. The rapidity coverage is
3.3$<$y$<$4.3 for the $\rho$, at low p$_{T}$, compared to 3$<$y$<$4
for the J/$\psi$.  The acceptance, about 1\% or smaller relative to
4$\pi$ in the whole low-mass region~\cite{michele:2005}, increases
with p$_{T}$ by a factor of 3 from 0 to 2~GeV/c for the $\rho$ and,
integrated over p$_{T}$, increases with mass by factors of 3-6 from
0.4 to 0.8 GeV/c$^{2}$, depending on the input p$_{T}$ spectrum. Finally,
the selective dimuon trigger and the fast readout speed of the vertex
tracker allow the experiment to run at very high luminosities, leading
to an unprecedented level of statistics for low-mass lepton pairs.

The results reported in this Letter were obtained from the analysis of
data taken in 2003 with a 158 AGeV indium beam, incident on a
segmented indium target of seven disks with a total of 18\% (In-In)
interaction length. At an average beam intensity of 5$\cdot$10$^{7}$
ions per 5~s burst, about 3$\cdot$10$^{12}$ ions were delivered to the
experiment.  A total of 230 million dimuon triggers were recorded on
tape, collected with two field settings of the muon-spectrometer
magnet; the data reported here correspond to the lower of the two
settings.

The data reconstruction starts with the muon-spectrometer
tracks. Next, pattern recognition and tracking in the vertex telescope
are done; the interaction vertex in the target is reconstructed with a
resolution of $\sim$200 $\mu$m for the z-coordinate and 10-20 $\mu$m
in the transverse plane. Only events with one vertex are kept;
interaction pileup and reinteractions of secondaries and fragments
are rejected. Finally, each muon-spectrometer track is extrapolated to
the vertex region and matched to the tracks from the vertex
telescope. The matching is done using the square of a weighted
distance ($\chi^{2}$), in a space of angles and inverse momenta,
between the two tracks~\cite{Ruben:2005qm}. Correct matches are
statistically distinguished from fake matches (associations of muons
to non-muon vertex tracks) by the very different shape of the matching
$\chi^{2}$ distributions. Matched tracks are kept if their $\chi^{2}$
is below a certain threshold. Pair reconstruction efficiencies
vary by $<$10\% both vs. mass and vs. centrality.

The combinatorial background of uncorrelated muon pairs mainly
originating from $\pi$ and K decays is determined using a {\it
mixed-event technique}~\cite{Ruben:2005qm}. Two single muons from
different like-sign dimuon triggers are combined into muon pairs in
such a way as to accurately account for details of the acceptance and
trigger conditions. The quality of the mixed-event technique can be
judged by comparing the like-sign distributions generated from mixed
events with the measured like-sign distributions. It is remarkable
that the two agree to within $\sim$1\% over a dynamic range of 4
orders of magnitude in the steeply falling mass
spectrum~\cite{Ruben:2005qm}. After subtraction of the combinatorial
background, the remaining opposite-sign pairs still contain ``signal''
fake matches, a contribution which is only 7\% of the combinatorial
background level. It has been determined in the present analysis by an
overlay Monte Carlo method. We have verified that an event-mixing
technique gives the same results, both in shape and in yield, within
better than~5\%. More details on the experimental apparatus and data
analysis will be given in a forthcoming extended paper; for now
see~\cite{Ruben:2005qm,Andre:2006}

\begin{figure}[t!]
\begin{center}
\includegraphics*[width=0.4\textwidth,clip=, bb = 6 12 560 665]{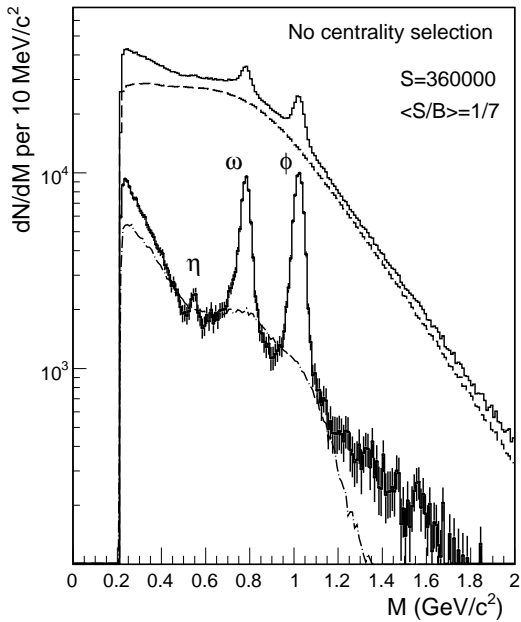}
\vspace*{-0.3cm}
\caption{Mass spectra of the opposite-sign dimuons (upper
histogram), combinatorial background (dashed), signal fake
matches (dashed-dotted), and resulting signal (histogram with error bars).}
\label{fig1}
\end{center}
\vspace*{-0.63cm}
\end{figure}

Fig.~\ref{fig1} shows the opposite-sign, background and signal dimuon
mass spectra, integrated over all collision centralities. After
subtracting the combinatorial background and the signal fake matches,
the resulting net spectrum contains about 360\,000 muon pairs in the
mass range of the figure, roughly 50\% of the total available
statistics.  The average charged-particle multiplicity density
measured by the vertex tracker is $dN_{ch}/d\eta$ =120, the average
signal-to-background ratio is 1/7. For the first time in nuclear
collisions, the vector mesons $\omega$ and $\phi$ are completely
resolved in the dilepton channel; even the
$\eta$$\rightarrow$$\mu$$\mu$ decay is seen. The mass resolution at
the $\omega$ is 20 MeV/c$^{2}$. The subsequent analysis is done in
four classes of collision centrality defined through the
charged-particle multiplicity density: peripheral (4-30),
semiperipheral (30-110), semicentral (110-170) and central (170-240).
The signal-to-background ratios associated with the individual classes
are 2, 1/3, 1/8 and 1/11, respectively.

The peripheral data can be described by the expected electromagnetic
decays of the neutral mesons~\cite{Damjanovic:qm2005}. Muon pairs from
resonance ($\eta, \rho, \omega, \phi$) and Dalitz ($\eta,
\eta^{'},\omega$) decays were simulated using the improved hadron
decay generator GENESIS~\cite{genesis:2003} as the input and GEANT for
transport through the detectors. Four free parameters (apart from the
overall normalization) were used in the fit of this ``hadron decay
cocktail'' to the peripheral data: the cross section ratios
$\eta/\omega$, $\rho/\omega$ and $\phi/\omega$, and the level of D
meson pair decays; the ratio $\eta^{'}/\eta$ was kept fixed at
0.12~\cite{Agakichiev:mv,genesis:2003}. The fits were independently
done in 3 bins of dimuon transverse momentum: p$_{T}$$<$0.5,
0.5$<$p$_{T}$$<$1 and p$_{T}$$>$1~GeV/c.
%%%%%%%%%%%%%%%%% figure 1.....
%%%%%%%%%%%%%%%%% end figure 1 .....
The particle ratios obtained from the fits without p$_{T}$ selection,
corrected for acceptance and extrapolated to full phase space are
$\eta/\omega$=0.88$\pm$0.03, $\phi/\omega$=0.094$\pm$0.004 and
$\rho/\omega$=1.62$\pm$0.10. The errors quoted are purely statistical;
the systematic errors are estimated to be of order 10\%, dominated by
uncertainties of the branching ratios (using the much more accurate
value for $\omega\rightarrow e^{+}e^{-}$ instead of $\mu^{+}\mu^{-}$
with the assumption of $e\mu$ universality). The value for
$\eta/\omega$ agrees, within errors, with the literature average of
0.82$\pm$0.11 for p-p, p-Be~\cite{Agakichiev:mv}. The value for
$\phi/\omega$ is subject to some $\phi$-enhancement already in
peripheral nuclear collisions and can therefore not be directly
compared to p-p values. Within 10\% these two ratios do not depend on
the pair p$_{T}$. In contrast, the particle ratio $\rho/\omega$
monotonically decreases with p$_{T}$~\cite{Damjanovic:qm2005},
reaching 1.20$\pm$0.09 at p$_{T}$$>$1~GeV/c, where it agrees with the
literature average of 1.0$\pm$0.15 for p-p, p-Be~\cite{Agakichiev:mv}
within errors. This suggests that some $\pi\pi$ annihilation,
enhancing the yield of the low-p$_{T}$ $\rho$, contributes already in
peripheral collisions (see below). The level of charm decays obtained
from the fits is essentially determined by the measured yield in the
mass interval 1.2$<$M$<$1.4~GeV/c$^{2}$. The results altogether, and
in particular those on the p$_{T}$ independence of $\eta/\omega$ and
$\phi/\omega$, indicate that the acceptance of NA60 both in mass and
p$_{T}$, including the critical low-mass ($\eta$-Dalitz), low-p$_{T}$
region ($<$0.5~GeV/c), is reasonably well understood. Apart from the
sources discussed, no further input was required to describe the
peripheral data.

%%%%%%%%%%%%%%%%%%figure 2
\begin{figure}[t!]
\begin{center}
\includegraphics*[width=0.4\textwidth,clip=, bb= 0 12 560 665]{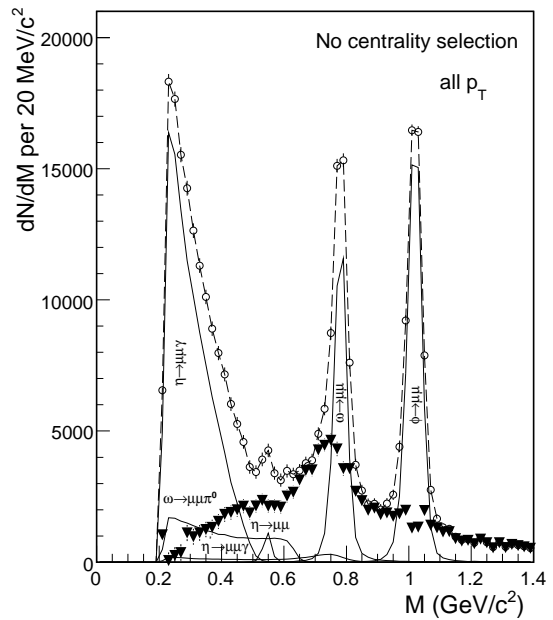}
\vspace*{-0.23cm}
\caption{Isolation of an excess above the electromagnetic decays
of a ``conservative'' hadron decay cocktail (see text). Total
data (open circles), individual cocktail sources (solid), difference
data (thick triangles), sum of cocktail sources and difference data
(dashed).}
   \label{fig2}
\end{center}
\vspace*{-0.6cm}
\end{figure}
%%%%%%%%%%%%%%%%% end of figure 2

\begin{figure}[t!]
\begin{center}
\includegraphics*[width=4.25cm, height=3.cm]{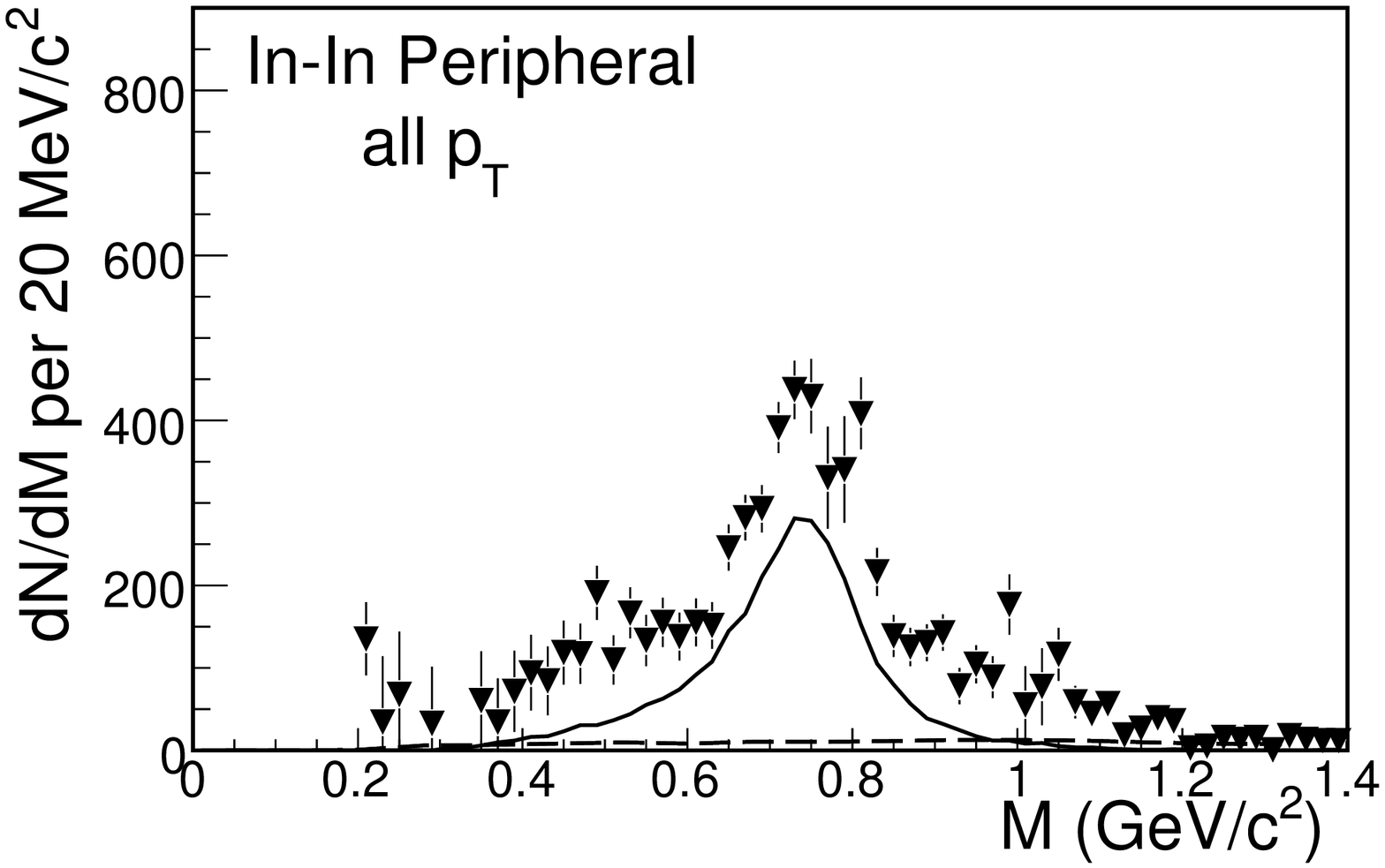}
\includegraphics*[width=4.25cm, height=3.cm]{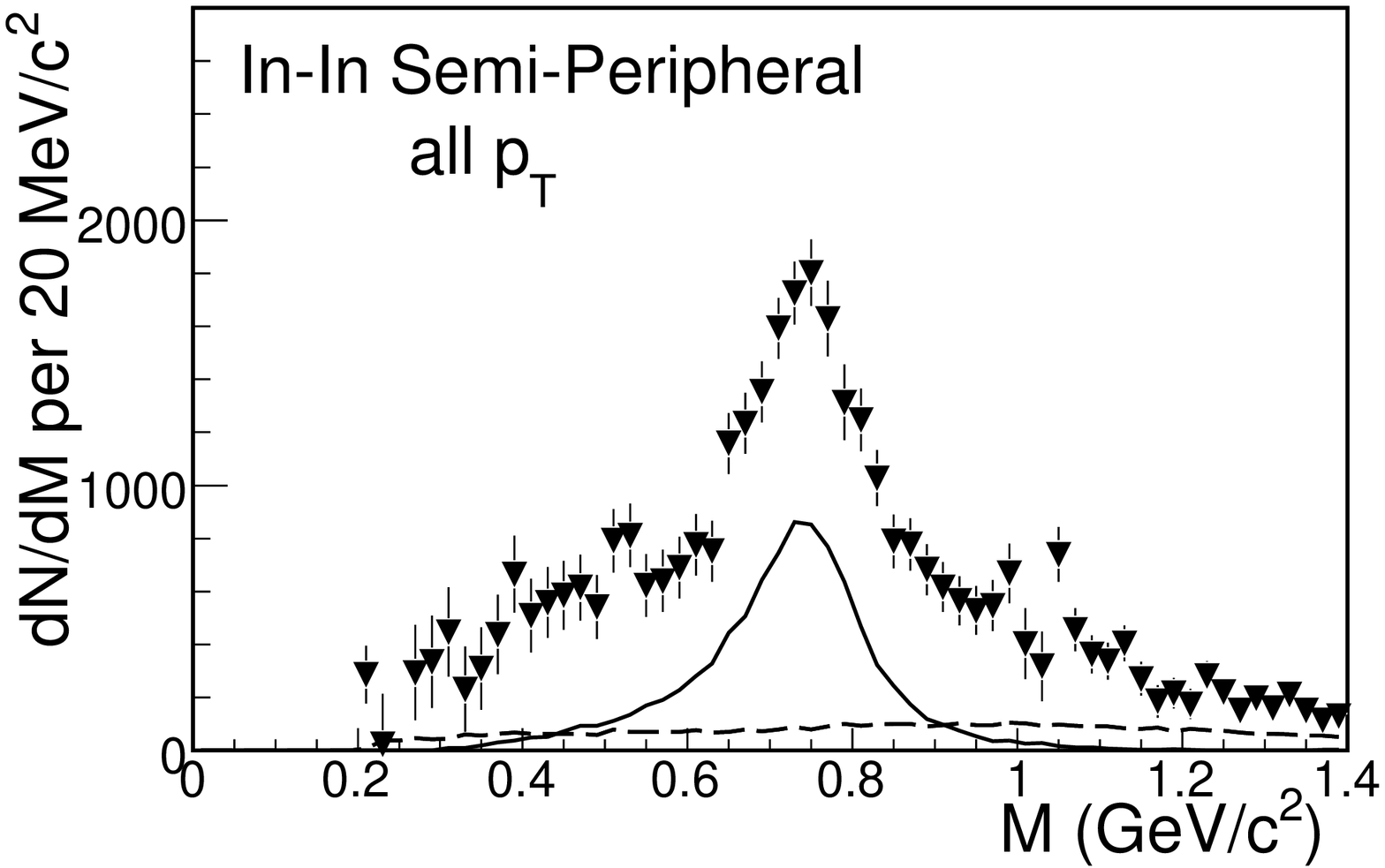}
\includegraphics*[width=4.25cm, height=3.cm]{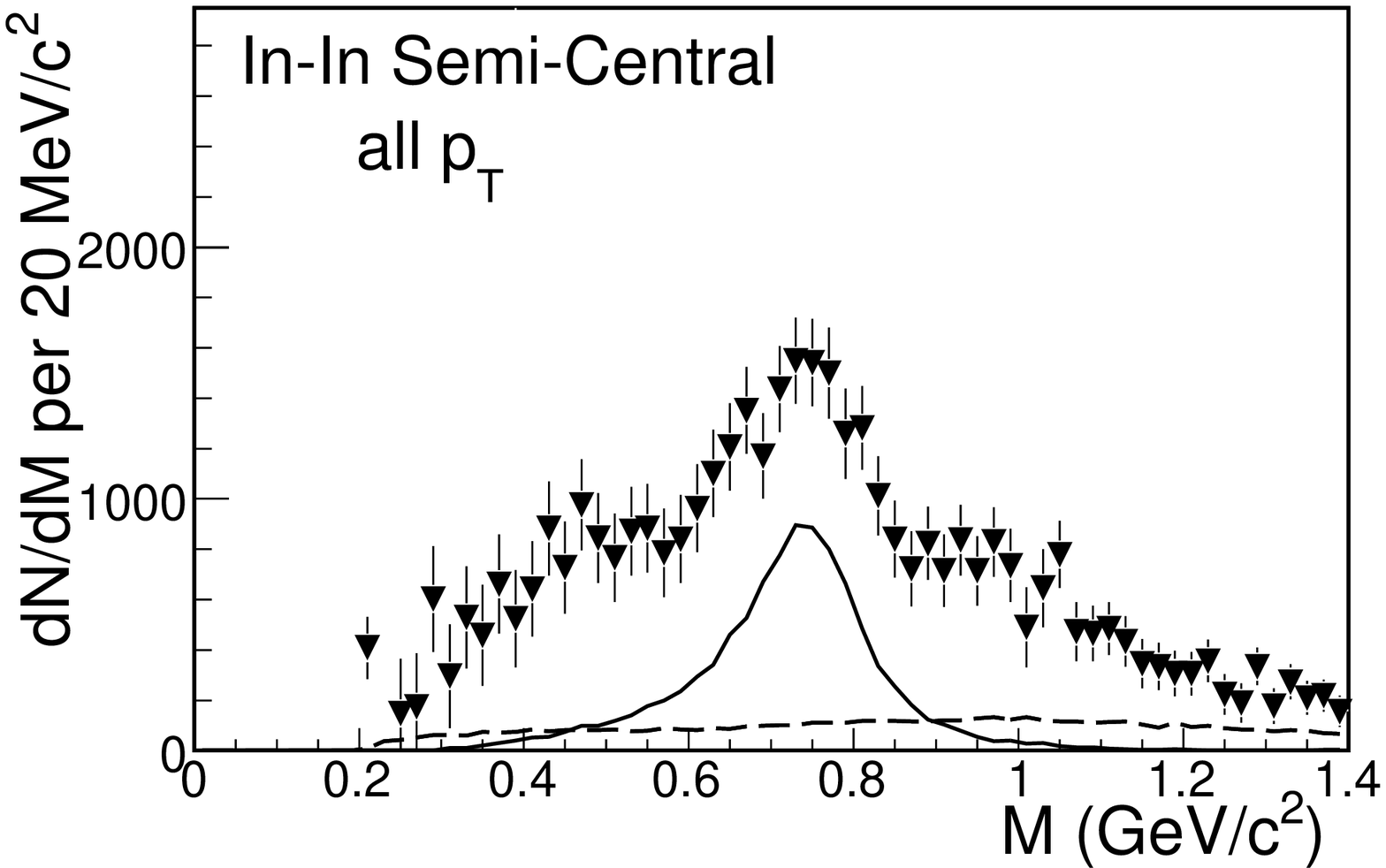}
\includegraphics*[width=4.25cm, height=3.cm]{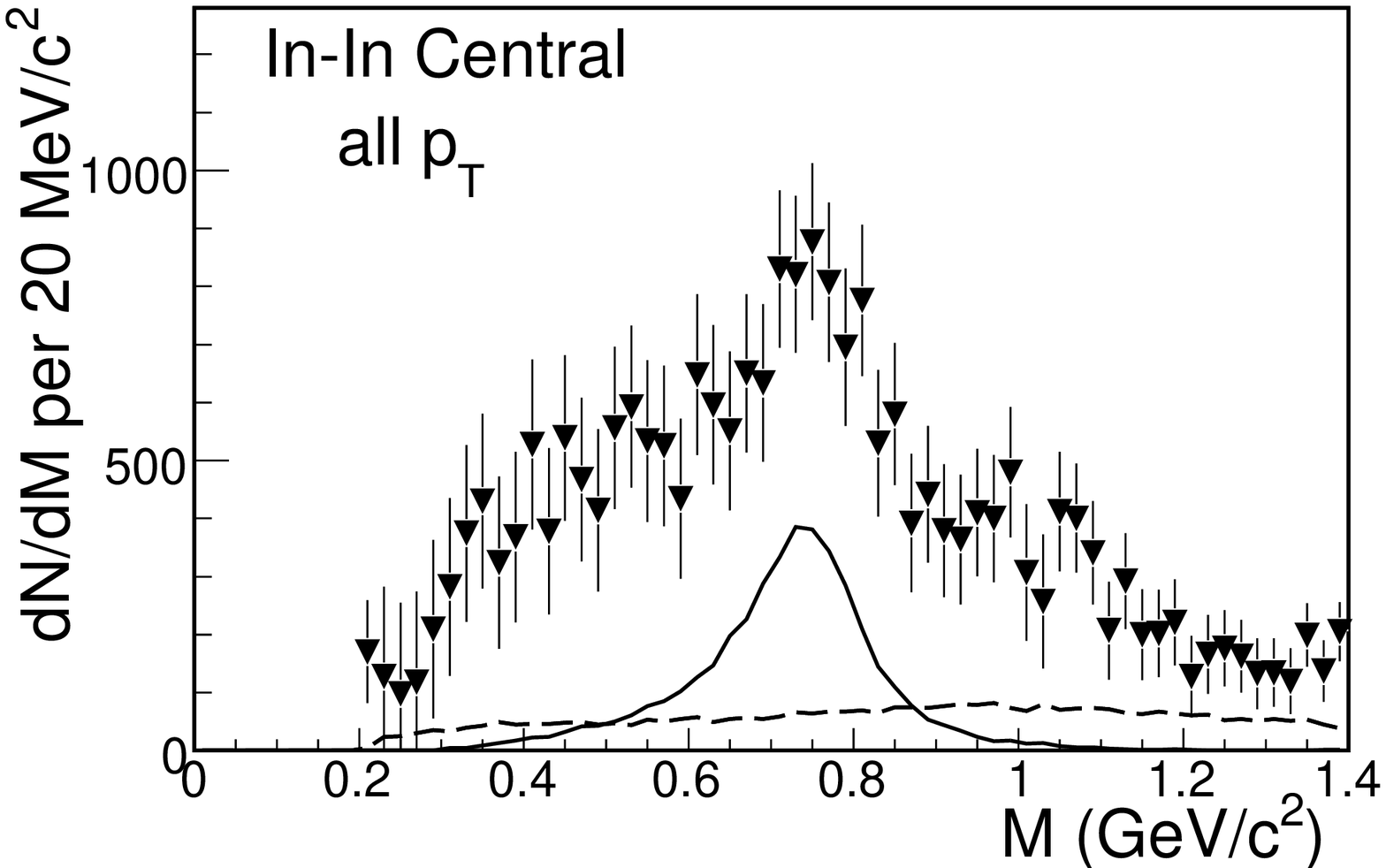}
\vspace*{-0.3cm}
   \caption{Excess mass spectra of dimuons. The cocktail $\rho$
   (solid) and the level of uncorrelated charm decays
   (dashed) are shown for comparison. The errors are purely
   statistical; see text for systematic errors.}
   \label{fig3}
\end{center}
\vspace*{-0.6cm}
\end{figure}

\begin{figure}[h!]
\begin{center}
\includegraphics*[width=0.4\textwidth]{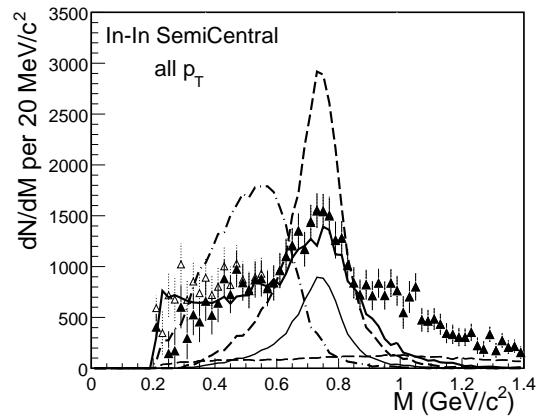}
\vspace*{-0.15cm}
\caption{Comparison of the excess mass spectrum for the semicentral
bin to model predictions, made for In-In at
$dN_{ch}/d\eta$=140. Cocktail $\rho$ (thin solid), unmodified $\rho$
(dashed), in-medium broadening $\rho$~\cite{Rapp:1995zy,Rapp:1999ej}
(thick solid), in-medium moving $\rho$ related to~\cite{Brown:kk,Brown:2001nh}
(dashed-dotted). The errors are purely statistical. The
systematic errors of the continuum are about 25\% (see text). The open
data points show the difference spectrum resulting
from a decrease of the $\eta$ yield by 10\% (which should also be
viewed as a systematic error).}
   \label{fig4}
\end{center}
\vspace*{-0.3cm}
\end{figure}

In the more central bins, the data can no longer be described on the
basis of the standard hadron decay cocktail alone, but is indicative
of the existence of an excess yield. Since the particle ratios are
expected to be different from the peripheral data, global fits to the
more central data are bound to bias both the extracted
cocktail parameters and an excess with {\it a priori unknown}
characteristics. We have therefore used a novel procedure, made
possible by the high data quality; it is illustrated in
Fig.~\ref{fig2}. The excess is {\it isolated} by subtracting the
cocktail, without the $\rho$, from the data. The cocktail is fixed,
separately for the major sources and in each centrality bin, by a
``conservative'' approach. The yields of the narrow vector mesons
$\omega$ and $\phi$ are fixed so as to get, after subtraction, a
{\it smooth} underlying continuum. For the $\eta$, an upper limit is
defined by ``saturating'' the measured data in the region close to 0.2
GeV/c$^{2}$; this implies the excess to vanish at very low mass, by
construction.  The $\eta$ resonance and $\omega$ Dalitz decays are
now bound as well; $\eta^{'}/\eta$ is fixed as before. The {\it
cocktail $\rho$} (only required in Figs.~\ref{fig3}~and~\ref{fig4},
for illustration purposes) is bound by the ratio $\rho/\omega$=1.2,
found at high p$_{T}$ ($>$1.6~GeV/c) for all centralities. The accuracy
in the determination of the $\omega$ and $\phi$ yields by this
subtraction procedure is on the level of 1-2\%, due to the remarkable
{\it local} sensitivity, and not much worse for the $\eta$. The
qualitative features of the resulting difference spectrum are robust
towards yield changes even on the level of 10\%, again because the
consequences of such changes are highly localized.

%%%%%%%%%%%%%%%%%%%%%% figure 3.....
%%%%%%%%%%%%%%%%% end of fugure 3

The excess mass spectra for all 4 multiplicity bins, resulting from
subtraction of the ``conservative'' hadron decay cocktail from the
measured data, are shown in Fig.~\ref{fig3}. The cocktail $\rho$ and
the level of charm decays, found in the 3 upper centrality bins to be
about 1/3 of the measured yield in the mass interval 1.2$<$M$<$1.4
GeV/c$^{2}$~\cite{Ruben:2005qm}, are shown for comparison. The
qualitative features of the spectra are striking: a peaked structure
is seen in all cases, broadening strongly with centrality, but
remaining essentially centered around the position of the nominal
$\rho$ pole.  At the same time, the total yield increases relative to
the cocktail $\rho$, their ratio reaching values above 4 for
M$<$0.9~GeV/c$^{2}$ in the most central bin. Such values are
consistent with the results found by CERES~\cite{Agakichiev:1997au},
if the latter are also referred to the cocktail $\rho$ and rescaled
according to the different multiplicity density. The errors shown are
purely statistical. The dominant sources of systematic errors are
connected to the uncertainties in the levels of the combinatorial
background (1\%) and fake matches (5\%). On the basis of these values
and the signal-to-background ratios, the systematic errors in the
broad continuum region 0.4$<$M$<$0.6 and 0.8$<$M$<$1.0~GeV/c$^{2}$ are
estimated to be about 3\%, 12\%, 25\% and 25\% in the 4 centrality
bins, from peripheral to central. Uncertainties associated with the
hadron decay cocktail and its subtraction including branching ratios,
transition form factors and helicity distributions have been estimated
to be around 15\%. Since the background causing the dominant errors is
essentially flat, the $\rho$-like structure above the continuum is
much more robust.

The qualitative features of the mass spectra in~Fig.~\ref{fig3} are
consistent with an interpretation of the excess as dominantly due to
$\pi\pi$ annihilation. Among the many different theoretical
predictions for the properties of the intermediate $\rho$ mentioned in
the introduction, only two have been brought to a level suitable for a
quantitative comparison to the data: the broadening scenario
of~\cite{Rapp:1995zy,Rapp:1999ej} and the moving-mass scenario related
to~\cite{Brown:kk,Brown:2001nh}. Both are evaluated for In-In at
$dN_{ch}/d\eta$=140 within the same fireball evolution, taking
explicit account of temperature as well as of baryon
%%%%%%%%%%%%%%%%%%%%%% figure 4.....
%%%%%%%%%%%%%%%%%%% end of figure 4
density~\cite{rapp:nn23}. In Fig.~\ref{fig4}, these predictions (as
well as the unmodified $\rho$) are confronted with the data for the
semicentral bin (same charged-particle rapidity density). Note that
the integrals of the theoretical spectra are normalized to the data in
the mass interval M$<$0.9~GeV/c$^{2}$. The unmodified $\rho$ is
clearly ruled out. The specific moving-mass scenario plotted here,
which fitted the CERES data~\cite{Rapp:1999ej,Agakichiev:1997au}, is
also ruled out, showing the much improved discrimination power of the
present data. The broadening scenario appears more realistic. However,
the nearly symmetrical broadening around the $\rho$ pole seen in the
data is not reproduced by this model. The remaining excess at M$>$0.9
GeV/c$^{2}$ may well be related to the prompt dimuon excess found by
NA60 in the intermediate mass region~\cite{Ruben:2005qm}. Processes
other than 2$\pi$, i.e. 4$\pi$ ... could possibly account for
the region M$>$0.9~GeV/c$^{2}$~\cite{gale:nn}.

The data shown in this Letter have not been corrected for the mass-
and p$_{T}$-dependent acceptance of the NA60 setup. The theoretical
calculations shown in Fig.~\ref{fig4} were therefore also propagated
through the acceptance filter to allow for fair comparison with the
data. It is interesting to note that, {\it by coincidence}, the
propagation of theoretical calculations based on a white spectral
function (e.g. $q\overline{q}$ annihilation~\cite{rapp:nn23}) yields a
mass spectrum flat within 10\% up to about 1~GeV/c$^{2}$, 
without any bump structure. In other words, the always existing steep
rise of the theoretical input at low masses, due to the photon
propagator and a Boltzmann-like
factor~\cite{Rapp:1995zy,Rapp:1999ej,Brown:kk,Brown:2001nh}, is just
about compensated by the falling acceptance in this region as long as
no p$_{T}$ cut is applied.  The data and model predictions shown in
Figs.~\ref{fig3}~and~\ref{fig4} can therefore be interpreted as
spectral functions of the $\rho$, averaged over momenta and the
complete space-time evolution of the fireball.  The flat part of the
measured spectra may thus reflect the early history close to the QCD
boundary with a nearly divergent width, while the narrow peak on top
may just be due to the late part close to the thermal freeze-out,
approaching the nominal width.

%%%%%% Conclusion
We conclude, in quite general terms and independently of any
comparison to theoretical modeling, that the $\rho$ primarily broadens
in In-In collisions, but does not show any noticeable shift in mass.
More detailed work including precise p$_{T}$ dependences is under way
to consolidate these findings. 

%%%%%% Acknowledgments
%\begin{acknowledgments}
%This work was supported by the ....
%\end{acknowledgments}

% \clearpage
\end{document}